\documentclass[conference]{IEEEtran}
\IEEEoverridecommandlockouts
\usepackage[utf8]{inputenc}
\usepackage{cite}
\usepackage{amsmath,amssymb,amsfonts}
\usepackage{amsthm}
\usepackage{algorithm}
\usepackage{algpseudocode}
\usepackage{tablefootnote}
\usepackage{graphicx}
\usepackage{textcomp}
\usepackage{tikz}
\usepackage{float}
\usepackage{xcolor}
\usepackage{multirow}
\usepackage{subfigure}
\usepackage{adjustbox}
\usepackage{stfloats}
\usepackage{url}
\usepackage{balance}
\usepackage{epstopdf}
\def\BibTeX{{\rm B\kern-.05em{\sc i\kern-.025em b}\kern-.08em T\kern-.1667em\lower.7ex\hbox{E}\kern-.125emX}}
\def\imagunit{\mathsf{j}} % Imaginary number

\begin{document}

\title{{mmWave Coverage Extension Using Reconfigurable Intelligent Surfaces in Indoor Dense Spaces }  }

\author{

 \IEEEauthorblockN{Zhenyu Li$^*$, Ozan Alp Topal$^*$, {\"O}zlem Tu\u{g}fe Demir$^\dagger$, Emil Bj{\"o}rnson$^*$, Cicek Cavdar$^*$}
\IEEEauthorblockA{ {$^*$Department of Computer Science, KTH Royal Institute of Technology, Kista, Sweden}
\\ {$^\dagger$Department of Electrical-Electronics Engineering, TOBB University of Economics and Technology, Ankara, Turkey
		} \\
		{Email: zhenyuli@kth.se, oatopal@kth.se, ozlemtugfedemir@etu.edu.tr, emilbjo@kth.se, cavdar@kth.se }
}
}

\maketitle

\begin{abstract}
In this work, we consider the deployment of reconfigurable intelligent surfaces (RISs)  to extend the coverage of a millimeter-wave (mmWave) network in indoor dense spaces. We first integrate RIS into ray-tracing simulations to realistically capture the propagation characteristics, then formulate a non-convex optimization problem that minimizes the number of RISs under rate constraints. We propose a feasible point pursuit and successive convex approximation-based algorithm, which solves the problem by jointly selecting the RIS locations, optimizing the RIS phase-shifts, and allocating time resources to user equipments (UEs).  The numerical results demonstrate substantial coverage extension by using at least four RISs, and a data rate of 130\,Mbit/s is guaranteed for UEs in the considered area of an airplane cabin. 
\end{abstract}


\section{Introduction}
 The wireless network demand in 6G mobile networks proliferates, especially for indoor usage \cite{ericsson} thanks to new applications such as virtual reality and ultra-high definition video streaming. As it is already in use in 5G networks, millimeter-wave (mmWave) bands can meet unprecedented data rate requirements thanks to the available wide spectrum.  Compared to the sub-6\,GHz signals, the mmWave signals are more sensitive to the blockage induced by the objects and humans within the system (e.g., $30$\,dB loss by the human body \cite{itu}) and have negligible diffraction. % Therefore, mmWave signal propagation has been extensively studied for varying indoor applications, such as office and factory \cite{Rappaport_mmWave, factory}. 
There is a recent interest to use mmWave networks in public transportation vehicles (e.g., airplanes, high-speed trains) to support the aforementioned new applications. For example, the mmWave propagation has been measured in airplanes \cite{cabin_measure_60}. Due to the smaller space, many blocking objects, and almost static user equipments (UEs), we call these environments indoor dense space (IDS). As shown in \cite{icc_22}, the blockage problem becomes even more challenging in these environments compared to the other indoor scenarios. One way to ensure coverage for all passengers is increasing the number of base stations (BSs) or using higher transmission power. However, due to the high cabling and energy cost, these solutions are not feasible in IDSs.    

Recently, reconfigurable intelligent surface (RIS) technology has emerged as a way to control the wireless communication channel. By deploying RISs that can adapt their reflection characteristics, we can passively redirect signals to the UEs in the outage zone and enhance the coverage area \cite{zhang2021reconfigurable}. The prototype described in \cite{prototype} demonstrates a gain of $26$\,dB in an  indoor space, where  the transmitter and receiver are separated by a 30\,cm thick concrete wall. The signal propagation with RIS has also been modeled by ray-tracing (RT) simulations in mmWave bands in outdoor \cite{outdoor_RIS} and indoor \cite{indoor_RIS} scenarios, and substantial performance gains were observed in both cases.  
In light of this, RIS seems a natural integration to the mmWave networks for IDSs. The blockage caused by dense objects can be overcome with a low deployment cost for the service providers by using RISs. Small-sized surfaces (e.g., for a $256$-element square RIS with quarter-wavelength spacing, an area of $4\times 4\,\textrm{cm}$ at $30$\,GHz carrier frequency) can be distributed among the medium to provide the best gains to all UEs. However, the deployment cost of the RISs makes it desirable to deploy a minimum number of RIS at the best possible locations while satisfying the quality-of-service requirements of the UEs.  This work investigates the potential performance gains obtained by deploying RISs in IDS mmWave networks. The proposed system  assigns UEs orthogonal time resources and configures phase-shifts of all deployed RISs to jointly serve a specific UE at a given time resource. The main contributions are as follows:
\vspace{-2mm}
\begin{itemize}
	\item We model an RIS-assisted mmWave network in an exemplary IDS environment. We provide new guidelines for how to implement RIS in RT simulations to realistically determine the propagation characteristics.
	\item  We cast a novel deployment problem that jointly minimizes the number of deployed RISs in a given IDS and optimizes the phase-shifts to meet the given data rate requirement at each UE location.
	\item We transform the original non-convex problem into a manageable form to apply feasible point pursuit and successive convex approximation (FPP-SCA). At each iteration of the proposed algorithm,  a mixed-integer cone programming problem is solved by the branch-and-bound technique. Thanks to the proposed reformulation, the binary variables are the only factors that violate the convexity of the problem, and the optimal solution at each iteration is guaranteed. 
	\item The numerical results demonstrate that RISs can significantly improve the coverage of a mmWave network for IDS environments. Moreover, the configuration cost can be minimized according to the data rate requirements by deploying RISs more sparsely.
\end{itemize}

 %   \subsection{Related Work}
        
  %      Making the wireless network intelligent and distributed is one of the expectations in 6G. Intelligent Reflecting Surface (RIS) is a new technology that is considered to have the potential to fulfill such expectation, since RIS makes it possible to control the wireless environment \cite{renzo2019smart}. The idea of the RIS is by integrating a massive number of reflecting elements, whose electromagnetic properties are reconfigurable, to control the reflecting signal \cite{wu2019towards}. 
        
   %     One of the important applications of RIS is to extend the coverage of a communication system. By deploying RIS close to the dead zone, it can passively redirect signals to the user in the area, thus enhancing the link quality \cite{zhang2021reconfigurable}. Comparisons are made between RIS and the traditional decode and forward (DF) relaying in \cite{bjornson2019intelligent}, showing that RIS can beat the DF relay with reasonable size and relatively low cost. In \cite{li2021enhancing}, \cite{nemati2020ris}, RIS-assisted communication system is investigated in a simplified environment. And the hardware imperfection of RIS elements is considered in \cite{nemati2020ris}.
        
    %   In this paper, an RIS-assisted communication system in a more complex indoor dense space (IDS) is investigated by using Ray-tracing (RT) simulation. And hardware imperfection is neglected in this work and assumes the phase shift in each RIS element can be chosen continuously. 
        \vspace{-2mm}
\section{System Model}
     %   \vspace{-2mm}

We consider an IDS environment, where each UE is located in a single passenger seat, and one BS is deployed in the center of the environment. The BS is equipped with multiple antennas in the form of a uniform planar array (UPA), while each UE has a single-antenna receiver. We let $k \in \{1,2,\ldots,K\}$ denote the UE indices, where $K$ is the total number of UEs. In our optimization problem, we will determine the deployment with the minimal number of RISs selected from the prospective RIS locations inside the cabin. We let $l\in \{1,2,\ldots, L\}$ denote the indices of the possible RIS deployments, where $L$ is the maximum number of all possible RIS deployments. The  $M$  elements in each RIS are deployed as a UPA. 

We consider downlink communication and since an RIS can only be optimized for one UE at a time, the UEs are scheduled for different time periods, so there will be no inter-UE interference. At a given time instance, the desired signal strength, and, thus, the data rate, of a UE  can be maximized by adjusting the phase-shifts of all the RISs in the cabin in a constructive way. Hence, during a particular UE's service time, all the RISs update their phase reconfiguration to direct their reflected beams to the served UE. We want to find the minimum number of RISs needed to satisfy the given target data rate requirement for each UE. By doing that, we will also find the best locations for the RISs and how many are required to be employed.  In the following, we will utilize the binary variable $\alpha_l\in \{0,1\}$ to denote whether RIS $l$ is deployed or not.
\vspace{-1mm}
\subsection{Channel Modeling and Downlink Data Transmission}
\vspace{-1mm}

Let $\mathbf{h}_k\in \mathbb{C}^N$ be the direct channel from the BS to UE $k$, where $N$ is the number of antennas at the BS. Moreover, $\mathbf{G}_l\in \mathbb{C}^{M\times N}$ denotes the channel from the BS to RIS $l$ and $\mathbf{g}_{l,k}\in \mathbb{C}^{M}$ is the channel from RIS $l$ to UE $k$. Due to the static nature of IDS channels \cite{icc_22}, we consider the channels as deterministic and fixed, and  solve a deployment problem based on the channel responses obtained by the RT. Let $\boldsymbol{\phi}_{l,k} \in \mathbb{C}^M $ denote the vector with the phase-shifts of RIS $l$ configured for serving UE $k$. The $m$th entry of the phase-shift vector  $\boldsymbol{\phi}_{l,k}$ is $\phi_{l,k,m}=e^{-\imagunit \varphi_{l,k,m} }$, where $\varphi_{l,k,m}\in[0,2\pi)$ is the phase-shift introduced by RIS element $m$ of surface $l$ while UE $k$ is being served. After the joint optimization of the RIS deployment, time allocation, and RIS phase-shift, the phase-shifts will be fixed for each selected RIS and served UE since the channels are constant. 

The transmitted signal from the BS to UE $k$ is  $\mathbf{x}_k=\mathbf{w}_k \eta_k$, where $\eta_k\in\mathbb{C}$ is the  data symbol and $\mathbf{w}_k \in \mathbb{C}^{N}$ is the unit-norm precoding vector that is selected based on the channel responses and RIS phase-shifts. The transmit power for each UE is $\mathbb{E}\{|\eta_k|^2\}=P$. Defining the diagonal matrix $\overline{\mathbf{G}}_{l,k}\in \mathbb{C}^{M \times M}$ with the diagonal entries as the entries of the vector $\mathbf{g}_{l,k}$, the received signal at UE $k$ can be written as
\begin{equation} \label{eq:receivedsignal}
   \mathbf{y}_k = \bigg( \mathbf{h}_k + \sum\limits_{l=1}^{L} \alpha_l\underbrace{\mathbf{G}_l^T\overline{\mathbf{G}}_{l,k}}_{\triangleq\mathbf{H}_{l,k}} \boldsymbol{\phi}_{l,k} \bigg)^{T}\mathbf{w}_k \eta_k + n_k,
\end{equation}
where $n_k\sim \mathcal{CN}(0,N_0)$ is the independent receiver noise. As seen from \eqref{eq:receivedsignal}, there is a BS-RIS-UE channel contribution for each selected RIS $l$ for $\alpha_l=1$. In \eqref{eq:receivedsignal}, we also have defined $\mathbf{H}_{l,k}=\mathbf{G}_l^T\overline{\mathbf{G}}_{l,k}\in \mathbb{C}^{N \times M}$.  Since there is no interference, the optimal precoding strategy for the BS is maximum ratio transmission (MRT), where 
\begin{equation}
   \mathbf{w}_k = \frac{\left(\mathbf{h}_k + \sum\limits_{l=1}^{L}\alpha_l\mathbf{H}_{l,k} \boldsymbol{\phi}_{l,k}\right)^* }{\left \Vert \mathbf{h}_k + \sum\limits_{l=1}^{L}\alpha_l\mathbf{H}_{l,k} \boldsymbol{\phi}_{l,k} \right\Vert}
\end{equation}
 is the conjugate of the overall effective channel normalized by its norm.  The resulting achievable data rate of UE $k$ is
 \begin{align}
     R_k = \tau_{k}B\log_2\left(1+\mathrm{SNR}_k\right) \quad \textrm{bit/s},
 \end{align}
 where $B$ is the communication bandwidth in Hz and $0\leq\tau_k\leq 1$ is the portion of the time allocated to UE $k$. Hence, $\sum_{k=1}^K\tau_k\leq1$. The term inside the logarithm is the one plus signal-to-noise ratio (SNR) at UE $k$, computed as
 \begin{equation}
   \mathrm{SNR}_k = \frac{ \left \lVert \mathbf{h}_k + \sum\limits_{l=1}^{L}\alpha_l\mathbf{H}_{l,k} \boldsymbol{\phi}_{l,k} \right \rVert^2 P}{ B N_0},
\end{equation}
 where $N_0$ is the noise power spectral density in W/Hz. In the following section, we will introduce the RIS deployment problem to determine the minimum number of RISs, which provide a certain data rate to each UE in the cabin.
 
\section{Joint RIS Deployment, Time Allocation, and RIS Phase-Shift Optimization}

The main goal of this paper is to minimize the number of deployed RISs in the cabin while guaranteeing the data rate of UE $k$ is above the threshold $\overline{R}_k$\,bit/s, for $k=1,\ldots,K$. Recalling that $\phi_{l,k,m}$ denotes the $m$th entry of the RIS phase-shift vector $\boldsymbol{\phi}_{l,k}$, this RIS deployment optimization problem can be formulated as 

\begin{subequations} \label{eq:Optimization1}
\begin{align} 
\textbf{P1: } & \underset{  \left\{\alpha_l,\tau_{k}, \boldsymbol{\phi}_{l,k} \right \} }{\textrm{minimize}} \quad\sum\limits_{l=1}^{L} \alpha_l \label{eq:objective}\\
& \textrm{subject to} \nonumber \\ &   \tau_k B\log_2\left(1+\frac{ \left \lVert \mathbf{h}_k + \sum\limits_{l=1}^{L}\alpha_l\mathbf{H}_{l,k} \boldsymbol{\phi}_{l,k} \right \rVert^2 P}{B N_0}\right) \nonumber \\
&\hspace{6mm}\geq \overline{R}_k, \quad \forall k, \label{eq:constraint1} \\ 
&  \sum_{k=1}^{K} \tau_{k}\leq 1,  \label{eq:constraint2}\\
 &  |\mathbf{\phi}_{l,k,m}|  = 1, \quad \forall l,k,m, \label{eq:constraint3} \\ &   \alpha_l \in \{0,1\}, \quad \forall l, \label{eq:constraint4}
 \end{align}
\end{subequations}
where the objective function in \eqref{eq:objective} is the total number of deployed RISs selected from $L$ potential locations in terms of binary RIS deployment variables. The constraint in \eqref{eq:constraint1} guarantees that the minimum data rate requirement for each UE is satisfied.  The constraint in  \eqref{eq:constraint2} ensures that the summation of the time portions allocated to the UEs does not exceed one. The unit-modulus constraints for the entries of the RIS phase-shift vectors $\boldsymbol{\phi}_{l,k}$ are given in \eqref{eq:constraint3}. Finally, \eqref{eq:constraint4} enforces the RIS deployment variables to be binary. We notice that, even if there were no binary variables, the constraints in \eqref{eq:constraint1} and \eqref{eq:constraint3} would destroy the convexity of the problem. To obtain a solution to this problem,  we will apply SCA. We first introduce new optimization variables $\tilde{\tau}_k$ in place of $1/\tau_k$ in \eqref{eq:constraint1} to later express the corresponding constraints in exponential cone form, which is a convex form. Moreover, we relax the unit-modulus constraints in \eqref{eq:constraint3} and obtain the modified optimization problem as
\begin{subequations}
\begin{align} \label{eq:Optimization2}
\textbf{P2: }&\underset{  \left\{\alpha_l,\tau_{k}, \tilde{\tau}_{k}, \boldsymbol{\phi}_{l,k} \right \} }{\textrm{minimize}}  \quad \sum_{l=1}^L \alpha_l  \\
& \textrm{subject to} \nonumber \\
&  \log_2\left(1+\frac{ \left \lVert \mathbf{h}_k + \sum\limits_{l=1}^{L}\alpha_l\mathbf{H}_{l,k} \boldsymbol{\phi}_{l,k} \right \rVert^2 P}{B N_0}\right) \nonumber \\
&\hspace{6mm}-  \frac{\overline{R}_k\tilde{\tau}_k}{B} \geq 0, \quad \forall k, \label{eq:constraintb1} \\  &   |\phi_{l,k,m}|   \leq 1, \quad \forall l,k,m,  \label{eq:constraintb2}\\
& \left \Vert  \left [
\tau_k,\  \tilde{\tau}_k,  \ \sqrt{2}\right]^T\right\Vert \leq \tau_k + \tilde{\tau}_k , \quad \forall k,  \label{eq:constraintb3}\\
& \text{\eqref{eq:constraint2}, \eqref{eq:constraint4}},
\end{align}
 \end{subequations}
where we have introduced the second-order cone (SOC) constraints in \eqref{eq:constraintb3} to construct the relation between $\tilde{\tau}_k$ and $\tau_k$. Those convex constraints guarantee that $\tilde{\tau}_k\geq \frac{1}{\tau_k}$. The inequalities  do not modify the optimal time portions $\tau_k$ for the problem \textbf{P2}. To see this, suppose that the optimal solution is such that for at least one $\tilde{\tau}_k$, the respective SOC constraint is satisfied with strict inequality, which leads to $\tilde{\tau}_k>\frac{1}{\tau_k}$. In such a case, $\tilde{\tau}_k$ can be reduced to $\frac{1}{\tau_k}$ without violating the data rate constraints in \eqref{eq:constraintb1}.

We notice that the data rate constraints can be expressed as exponential cones by  introducing $d_k$ in place of the term $1+\mathrm{SNR}_k$, and we can rewrite the constraints in  \eqref{eq:constraintb1} equivalently as
\begin{align}
& d_k \geq 2^{\frac{\overline{R}_k\tilde{\tau}_k}{B}}, \quad \forall k, \label{eq:exponential_cone} \\
&   1+\frac{ \left \lVert \mathbf{h}_k + \sum\limits_{l=1}^{L}\alpha_l\mathbf{H}_{l,k} \boldsymbol{\phi}_{l,k} \right \rVert^2 P}{B N_0} \geq d_k, \quad \forall k, \label{eq:quadratic_nonconvex}
\end{align}
where the inequalities in \eqref{eq:quadratic_nonconvex} are the only non-convex constraints except for the binary variables. We define $\mathbf{H}_k = \left [\mathbf{H}_{1,k}, \ \cdots, \ \mathbf{H}_{L,k}\right]\in\mathbb{C}^{N\times LM}$ and the new optimization variables $\mathbf{z}_{k}=\left[\alpha_1\boldsymbol{\phi}_{1,k}^T, \ \cdots, \ \alpha_L\boldsymbol{\phi}_{L,k}^T\right]^T \in \mathbb{C}^{LM}$ to express \eqref{eq:quadratic_nonconvex} in quadratic form. Then, we can expand the norm square in \eqref{eq:quadratic_nonconvex} as
\begin{equation} \label{eq:norm-expand}\mathbf{z}_{k}^H\underbrace{\mathbf{H}_{k}^H\mathbf{H}_{k}}_{\triangleq\mathbf{A}_k}\mathbf{z}_{k} + 2\Re\left(\mathbf{z}_{k}^H\underbrace{\mathbf{H}_{k}^H\mathbf{h}_k}_{\triangleq\mathbf{b}_k}\right)+\underbrace{\mathbf{h}_k^H\mathbf{h}_k}_{\triangleq c_k}
\end{equation}
which leads to the constraints in \eqref{eq:quadratic_nonconvex} given as
\begin{align}
-\mathbf{z}_k^H\mathbf{A}_k\mathbf{z}_{k}-2\Re\left(\mathbf{z}_k^H\mathbf{b}_k\right) \leq c_k + \frac{BN_0}{P}(1-d_k) \label{eq:quadratic_nonconvex2}
\end{align}
which is still not convex. We will utilize the FPP-SCA algorithm to further handle the non-convexity\cite{mehanna2014feasible}. Since $\mathbf{A}_k$ is positive semi-definite, for any arbitrary vector $\boldsymbol{\zeta}_k$ it holds that
\begin{equation}
    \mathbf{z}_{k}^H(-\mathbf{A}_{k})\mathbf{z}_{k} \leq 2\Re\left(\boldsymbol{\zeta}_{k}^H\left(-\mathbf{A}_{k}\right)\mathbf{z}_{k}\right)-\boldsymbol{\zeta}_{k}^H(-\mathbf{A}_{k})\boldsymbol{\zeta}_{k}. \label{eq:convexified}
\end{equation}
In conventional SCA, the above affine approximation is used in the neighborhood of the solution found in the previous iteration. Inserting this approximation in place of $\mathbf{z}_{k}^H(-\mathbf{A}_{k})\mathbf{z}_{k}$ in \eqref{eq:quadratic_nonconvex2}, a mixed-integer programming problem is solved at each iteration. The respective optimization problem would be convex if there were no binary variables. Such a problem can be solved optimally at each iteration using branch-and-bound algorithms. In the FPP-SCA approach, extra slack variables $s_k\geq 0$ are included in the convexified constraints to solve the infeasibility issue, often observed in the initial iterations. By minimizing the summation of these non-negative slack variables, the respective inequalities are satisfied with equality at the solution. At each iteration $r$, inserting the previously obtained solution $\mathbf{z}_k^{(r-1)}$ in place of $\boldsymbol{\zeta}_k$ in \eqref{eq:convexified} and the slack variables $s_k\geq 0$ into \eqref{eq:quadratic_nonconvex2}, the following problem is solved optimally:
\begin{subequations}
	\begin{align}
        \textbf{P3: }&\underset{  \left\{\alpha_l,\tau_{k}, \tilde{\tau}_{k}, \mathbf{z}_{k}, d_k, s_k \right \} }{\textrm{minimize}}\quad \sum_{l=1}^L \alpha_l  + \Omega\sum_{k=1}^K s_k\\
        & \textrm{subject to} \nonumber \\
        & -2\Re\left(\left(\mathbf{z}_{k}^{(r-1)}\right)^H\mathbf{A}_{k}\mathbf{z}_{k}\right)+\left(\mathbf{z}_{k}^{(r-1)}\right)^H\mathbf{A}_{k}\mathbf{z}_{k}^{(r-1)} \nonumber\\
        &\hspace{6mm}-2\Re\left(\mathbf{z}_{k}^H\mathbf{b}_{k}\right)\leq s_{k}+c_{k}+\frac{BN_0}{P}(1-d_k), \quad\forall k, \\ 
        & 
        s_k \geq 0, \quad \forall k, \\   
        & 
        |z_{k,l,m}| \leq \alpha_l, \forall k, l, m, \\    
        & \text{\eqref{eq:constraint2}, \eqref{eq:constraint4}, \eqref{eq:constraintb3}, \eqref{eq:exponential_cone}, }
 \end{align}
\end{subequations}
where we have added a penalty term with $\Omega>0$ to the objective function to force the slack variables $s_k$ to zero as the iterations evolve. Furthermore, we have expressed the relation between $\alpha_l$ and the corresponding entries of $\textbf{z}_k$, which are $z_{k,l,m}$.  The steps of the FPP-SCA algorithm are outlined in Algorithm~\ref{alg:fpp}. Based on our empirical observations, we normalize the entries of the vectors $\mathbf{z}_k$ to push the solution onto the unit circle for each selected RIS with $\alpha_l=1$ in Line 5 of the algorithm for faster convergence. Omitting this normalization, when all $s_k$ become zero, any feasible solution to the problem \textbf{P3} will also be a feasible solution to the original problem $\textbf{P2}$ from the relation in \eqref{eq:convexified}. Since the solution improves or remains the same at each iteration and the objective function is bounded, the FPP-SCA algorithm converges. 

\begin{algorithm}
	\caption{FPP-SCA for RIS Deployment } \label{alg:fpp}
	\begin{algorithmic}[1]
		\State {\bf Initialization:} Initialize $\mathbf{z}_k^{(0)}$ randomly while keeping $\left|z_{k,l,m}^{(0)}\right| = 1, \forall k,l,m$. Set the penalty coefficient $\Omega = 100$. Set the iteration counter to $r=0$. Set the solution accuracy to $\epsilon>0$.
		\While{$\sum_{k=1}^K\left\Vert \mathbf{z}_k^{(r)}-\mathbf{z}_k^{(r-1)}\right\Vert^2>\epsilon$}
		    \State $r \gets r+1$
		    \State Solve problem \textbf{P3} with mixed integer programming (MIP) solver, and set $\textbf{z}_k^{(r)}, \forall k$ to its optimum solution
%    		\State Set $\textbf{z}_k^{(r)}, \forall k$ to the optimal solution of the problem \textbf{P3}  
    	    \State $z_{k,l,m}^{(r)} \gets z_{k,l,m}^{(r)}\Big/\left\vert z_{k,l,m}^{(r)}\right\vert, \quad \forall k, l, m$
		\EndWhile
		\State {\bf Output:} The selected RISs, i.e., $\alpha_l^{(r)}$, and their phase-shifts $\mathbf{z}_k^{(r)}$, and allocated time portions $\tau_k^{(r)}$ 
	    \end{algorithmic}
\end{algorithm}

\section{RIS Implementation in Ray-Tracing}
A commercial RT tool, Wireless Insite\footnote{Wireless InSite, available at: \url{http://www.remcom.com/wireless-insite}}, is used to accurately determine channel properties in complex environments \cite{yun2015ray} by using the shooting-and-bouncing ray (SBR) approach.
   % RT-based radio wave propagation modeling can accurately determine channel properties in complex environments \cite{yun2015ray}, thus it is well suited for determining the channels in mmWave bands because those are particularly sensitive to the geometry of the environment and objects within \cite{series2012propagation}. 
   Since the UEs are static in the IDS, we can obtain channel coefficients for all BS-RIS-UE links considering possible RIS deployments with RT.

\begin{figure}
    \vspace{8mm}
        \begin{center}	
            \subfigure[]{
                \label{fig:IDS_overview}
             \includegraphics[trim={4mm 25mm 4mm 22mm},clip,width=0.5\linewidth]
            {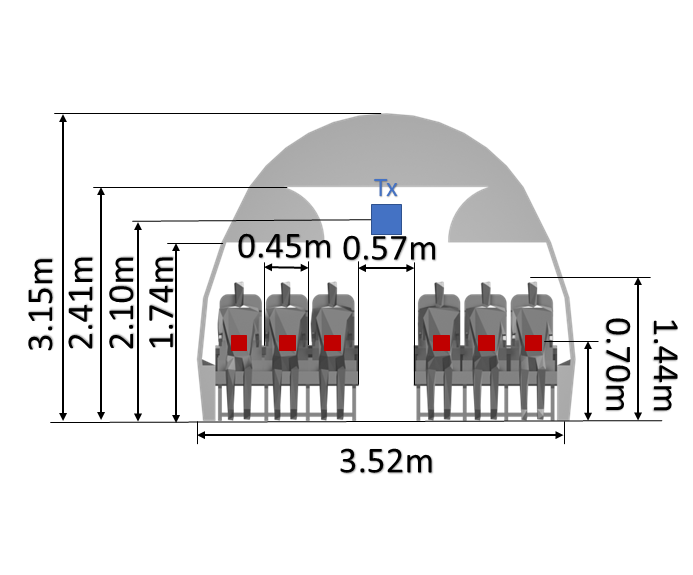} }
             
            \subfigure[]{
           \vspace{-0.5cm}     \label{fig:IDS_front}
              \includegraphics[trim={3mm 26mm 4mm 16mm},clip,width=0.45\linewidth]{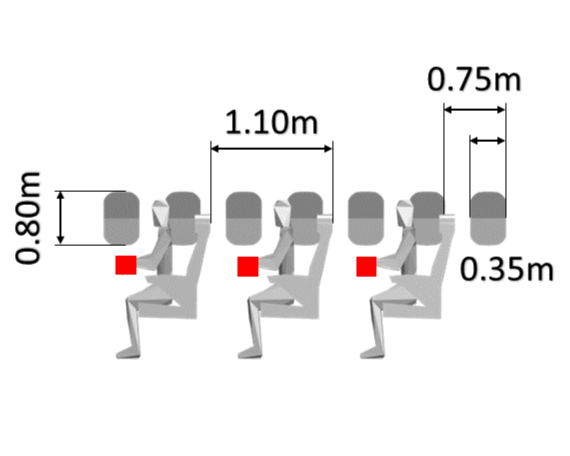} }
            \subfigure[]{
            \label{fig:IDS_side}
            \includegraphics[trim={3mm 4mm 4mm 4mm},clip,width=0.45\linewidth]{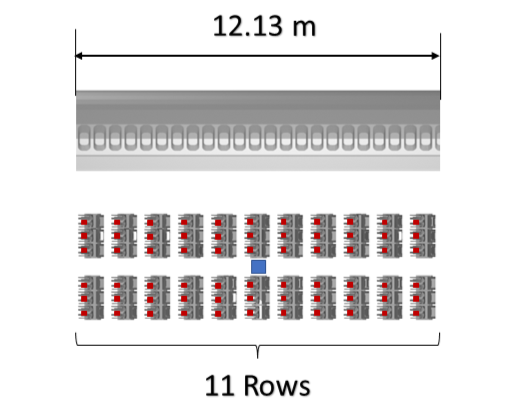} }
        \end{center}
        \vspace{-0.5cm}
        \caption{Illustration of the IDS RT environment and geometry. (a) Cabin front view, where the red box represents the UE receiver (Rx). (b) Cabin side view. (c) The overview of BS transmitter (Tx) and Rx locations.} 
        \label{fig:IDS_geometry}   
    \end{figure}

    The geometry of the considered airplane environment is shown in Fig.~\ref{fig:IDS_geometry}. To consider an area where a single BS can deliver high data rates, 11 rows that are fully seated with 66 passengers are considered. The BS is placed in the middle of the cabin. For each passenger, a receiving node is placed to represent the UE held by the passenger. %To obtain realistic simulation results, the materials of the different objects in the cabin are selected based on a real aircraft environment. 
    The material of the cabin shell is Acrylonitrile Butadiene Styrene (ABS), which is widely used in aircrafts~\cite{cabin_measure_60}. Windows and passenger seats are set to glass and nylon, respectively. Passengers are modeled by the same human skin model as in \cite{felbecker2008incabin}. The dielectric properties of the materials we consider are given in Table~\ref{tab:simulation_dielectric}.

    \begin{figure}
        \begin{center}
            \includegraphics[trim={3mm 3mm 4mm 2mm},clip,width=\linewidth]{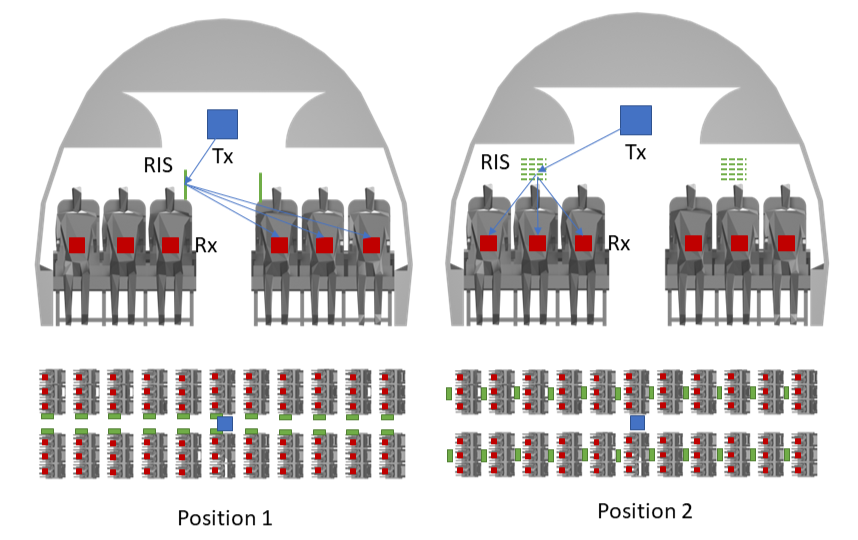}
        \end{center}
    \vspace{-4mm}
    \caption{Illustration of two candidate positions in the preliminary test.}
\label{fig:IDS_candidate_position}
    \end{figure}

     RIS is generally not  an integrated component in RT simulation platforms.  To realistically  mimic the behaviour of the RIS, we use a transceiver antenna array at the RIS locations to obtain the channel impulse responses (CIRs) of the BS-RIS and RIS-UE links \cite{indoor_RIS}. To realistically model RISs, we followed the following steps. First, an RIS with $256$ elements is represented by a $16 \times 16$ isotropic antenna array. The horizontal and vertical distances between the elements in the RIS are set to $\lambda/4$, where $\lambda$ is the wavelength of the carrier signal. Later, a perfect wave absorber with the same size as the transceiver antenna array is added 1\,cm behind the array since the RIS is not penetrable, and only one side of RIS reflects. We use the term reflective side to indicate the side without a perfect wave absorber.  The CIRs obtained from the transceiver-BS link and transceiver-UE link are regarded as $\mathbf{G}_l$ and $\mathbf{g}_{l,k}$, respectively. Since isotropic antennas are considered to simulate the RIS elements, the channel gains from/to RISs are scaled by the area  $(\lambda/4)^2$ of RIS elements divided by the effective area $\lambda^2/(4\pi)$ of the isotropic antenna.

\begin{figure}[t!]
    \vspace{2mm}
        \centering
        \includegraphics[trim={4mm 0mm 0mm 2mm},clip,width=0.85\linewidth]{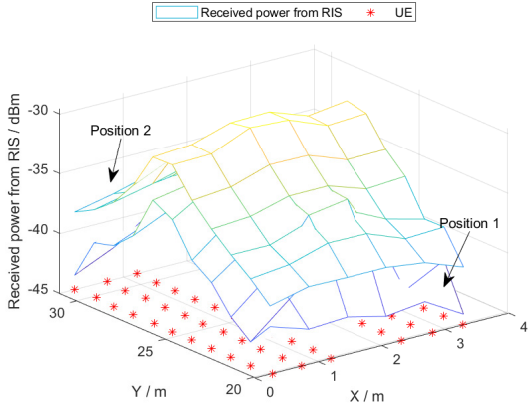}
        \vspace{-3mm}
        \caption{Preliminary test of the two candidate positions.}   \label{fig:RIS_position_compare}
          \vspace{-4mm}
    \end{figure}

    There are several different positions on each row where an RIS can potentially be deployed. A preliminary test was designed to compare the received signal power gains by RISs at different places. Two candidate positions were chosen as illustrated in Fig.~\ref{fig:IDS_candidate_position}. In both positions, RISs are perpendicular to the floor. The heights of the RIS centers are set to 1.7\,m. In candidate position 1, the RIS is placed in the corridor area close to the seat with the reflective side facing towards the corridor. In candidate position 2, the RIS is placed above the middle seats on both sides of the corridor, with the reflective side facing toward the BS. Since we are interested in determining the advantageous candidate position, we compare the channel gains of the RISs with the perfect phase-shift configuration at RISs as in \cite{bjornson2019intelligent}.
    The BS was considered to have a single isotropic antenna in this test without loss of generality since our aim is to determine the preferred location for the RISs.  The detailed simulation settings are  listed in Table~\ref{tab:simulation_parameters}. In Fig.~\ref{fig:RIS_position_compare}, it is observed that the received power from the RIS located in position 2 is higher, which can be intuitively explained by the fact that RIS is expected to provide higher gain when it is very close to the UE or BS. Hence, position 2 is selected for the RISs in candidate locations in the remainder.

    \vspace{-1mm}
    \begin{table}[H]
        \centering
        \caption{Dielectric properties of materials at $28$\,GHz considered in the simulation.}
         \vspace{-2mm}\label{tab:simulation_dielectric}
        \begin{tabular}{|l|cc|c|}
        \hline
        \multicolumn{1}{|c|}{\multirow{2}{*}{Material}} & \multicolumn{2}{c|}{dielectric properties} & Thickness (cm) \\ \cline{2-4} 
        \multicolumn{1}{|c|}{}                          & \multicolumn{1}{c|}{$\epsilon$}    & $\sigma$    & n/a            \\ \hline
        Skin \cite{wu2015human}  & \multicolumn{1}{c|}{$19.3$} & $30.40$ & $0.1$  \\ \hline
        ABS \cite{singh2017investigations}  & \multicolumn{1}{c|}{$2.4$}  & $0.028$ & $0.3$  \\ \hline
        Nylon \cite{riddle2003complex} & \multicolumn{1}{c|}{$3.01$} & $0.03$  & $0.25$ \\ \hline
        Glass \cite{series2012propagation} & \multicolumn{1}{c|}{$6.27$} & $0.15$  & $0.3$  \\ \hline
        \end{tabular}
        \vspace{-2mm}
    \end{table}

    \begin{table}[H]
            \vspace{-4mm}
        \centering
        \caption{Simulation parameters for the RT.}
        \label{tab:simulation_parameters}
        \vspace{-2mm}
        \begin{tabular}{|l|c|}
        \hline
        Carrier frequency      & $28$\,GHz         \\ \hline
        Bandwidth              & $1$\,GHz          \\ \hline
        Antenna type           & Isotropic      \\ \hline
      %  Propagation mechanisms & 2 Reflections  \\ \hline
        Polarization           & V - V          \\ \hline
       % RIS size               & $16 \times 16$ \\ \hline
       % RIS element interval   & $\lambda / 4$  \\ \hline
        Transmit power               & $25$\,dBm          \\ \hline
        Number of BS antennas        & $64 $\\ \hline
        Antenna spacing    & $\lambda / 2$  \\ \hline
        \end{tabular}
    \end{table}

\section{Numerical Results}
    
\begin{figure}[t!]
        \vspace{2mm}
        \centering
        \includegraphics[width=0.9\linewidth]{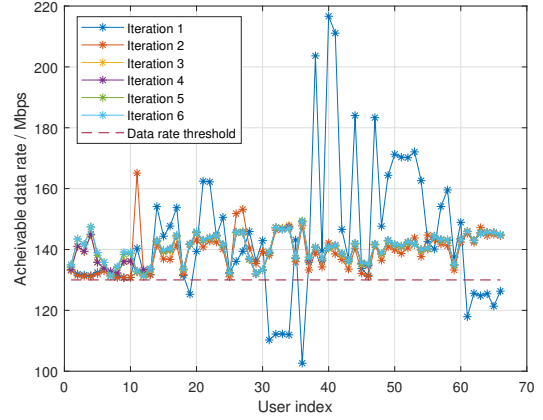}
        \vspace{-2mm}
        \caption{Achievable data rate of each iteration of the optimization algorithm under the $130$\,Mbit/s data rate threshold.}
        \label{fig:RIS_iteration_130_mpbs}
        \vspace{-4mm}
    \end{figure}

    We consider an $8 \times 8$ isotropic antenna array for the BS. The essential parameter values for this mmWave communication setup are given in Table~\ref{tab:simulation_parameters}. We solved \textbf{P2} using the proposed Algorithm 1 considering different data rate thresholds per UE with $64$ ($8 \times 8$) or $256$ ($16 \times 16$)  elements per RIS. Since there are two RISs per row, there are $L=11\cdot 2=22$ potential RIS locations.  %We set the initial value of the optimization variables $\mathbf{z}_k^{(0)}$ to the solution of the convex feasibility problem with the same data rate constraints for faster convergence.
    Fig.~\ref{fig:RIS_iteration_130_mpbs} shows the achievable data rate of each iteration when the data rate threshold is set to $130$\,Mbit/s considering 256 RIS elements. The first iteration of the algorithm fails to find the output that satisfies the data rate constraint, which shows that the corresponding slack variables $s_k$ are positive. In the next iteration, the algorithm finds a solution that satisfies the data rate requirement for all UEs, and it converges in around $6$ iterations. Similarly, for other data rate constraints, we have observed that the proposed algorithm converges in around $5$-$6$ iterations.  %Due to the introduction of the stack variable, the algorithm does not warn infeasibility during iterating.  
    
        \begin{figure}[t!]
        \centering
        \includegraphics[trim={0mm 0mm 0mm 0mm},clip,width=0.8\linewidth]{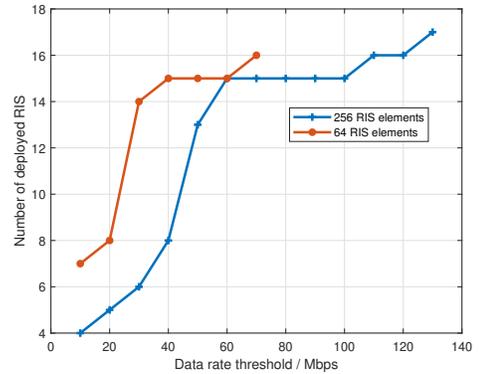}
        \vspace{-2mm}
        \caption{Number of required RISs to be deployed for each data rate threshold per UE. $256$ or $64$ elements are considered for each RIS.}
        \label{fig:RIS_number_vs_threshold}
        \vspace{-2mm}
    \end{figure}
    
    The data rate threshold per UE is varied from $10$\,Mbit/s to $140$\,Mbit/s with an increment of $10$\,Mbit/s. The number of deployed RISs for each data rate threshold, the objective of \textbf{P2}, is given in Fig.~\ref{fig:RIS_number_vs_threshold}. When the data rate threshold is low, RISs are mainly deployed to assist UEs that receive no signal from the BS (e.g., due to blockages). As the data rate requirement increases up to a certain level ($40$\,Mbit/s and $60$\,Mbit/s for $64$-element and $256$-element RIS, respectively), more RISs are deployed to increase the SNR levels at the UEs. In the range $40$-$60$\,Mbit/s for the $64$-element RIS and the range $60$-$100$\,Mbit/s, for the $256$-element RIS, no additional RIS is needed. This result shows that once we have extended the coverage by deploying RIS that supports all the far-away UEs, we can satisfy relatively high data rate requirements by optimizing only the time resources without the need for additional RIS deployments. When the data rate threshold becomes even higher, the nearby UEs that the BS could previously serve without requiring any RIS will also have difficulty satisfying the data rate threshold; then, more RISs are again deployed. By comparing the two RIS sizes, we notice that when having larger RIS sizes, we can both use fewer RISs and reach higher data rates in the upper tail of the curve. This comparison suggests that there is a trade-off between the number of elements in an RIS and the total number of RISs being deployed.
    
    %When the data rate threshold is between 60\,Mb/s to 100\,Mb/s, mainly UEs that are non-line-of-sight (NLOS) to the BS are in need of the assistance of RISs. Additionally, due to the direction of the reflective side of each RIS, the benefit obtained for UEs that are in the opposite direction of the reflective side is very limited. Thus, no more RISs being deployed when the data rate threshold is within this range. 

   % The algorithm is rerun for RIS with fewer elements. Showing in Fig.~\ref{fig:RIS_number_vs_threshold}, for the same data rate threshold, more RISs are deployed when each RIS contains fewer elements. 

\begin{figure}
        \centering
        \includegraphics[trim={0mm 4mm 0mm 0mm},clip,width=\linewidth]{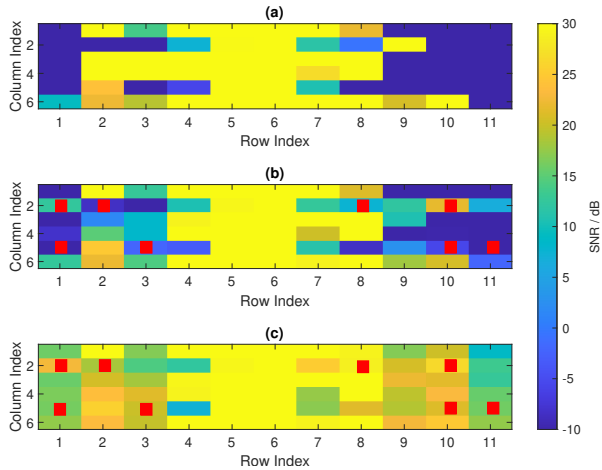}
        \vspace{-5mm}
        \caption{SNR of each UE. The red boxes indicate the deployed RISs.  The indices of the UEs and RISs refer to the geometric position illustrated in Figs.~\ref{fig:IDS_geometry} and \ref{fig:IDS_candidate_position}. (a) Without any RIS; (b) with the optimized RIS deployment and random RIS phase-shifts; (c) with the optimized RIS deployment and optimized phase-shifts under the 40\,Mbit/s data rate threshold per UE.} 
        \label{fig:RIS-free-optimal-random}
        \vspace{-3mm}
    \end{figure}

    To further analyze the benefit brought by the RISs, we compare three different cases in Fig.~\ref{fig:RIS-free-optimal-random}. In the first case, in Fig.~\ref{fig:RIS-free-optimal-random}(a), we present  the SNR of the UEs without any RIS deployment. Due to the simplicity of the propagation mechanisms considered during the RT simulation, some UEs have zero channel gain, indicating the necessity of RIS deployment to cover them. The coverage in this case is roughly 3 rows from both sides of the BS. In the second one, in Fig.~\ref{fig:RIS-free-optimal-random}(b), we show the SNR when optimized RIS deployment obtained from Algorithm~1 under the $40$\,Mbit/s data rate threshold per UE is used. However, instead of optimized RIS phase-shifts, randomly generated phase-shifts are induced by each RIS. We observe no substantial gains compared to the first case, illustrating the necessity of intelligent reconfiguration. In Fig.~\ref{fig:RIS-free-optimal-random}(c), we present the third case, where $8$ RISs are deployed at the optimized locations with the optimized phase-shifts to meet the $40$\,Mbit/s per UE threshold. In this case, the minimum SNR among the UEs is $8$\,dB, which is $18$\,dB higher than in the first and second cases. %Benefiting from the capability to control the wireless environment, RIS with the optimal phase shift can obviously improve the SNR condition.
     %UEs in this area obtain much higher SNR than those out of the area. In the $1^{st}$ row, $10^{th}$ row, and $11^{th}$ row, users with SNR worse than 0 dB can be observed. 
   % Due to the simplicity of the propagation mechanisms considered during RT simulation, illustrated in Fig.~\ref{fig:RIS_position_compare} (a), there are also users who receive no rays from the BS. Comparing to the results in Fig.~\ref{fig:RIS_position_compare} where all users have non-zero received power, one conclusion can be drawn is that there are always some RISs being deployed to serve UEs with heavy attenuation.
  % Fig.~\ref{fig:RIS-free-optimal-random} (b) and (c) illustrate the SNR of each UE with the optimal deployment output from the algorithm without and with optimal phase shift under the 40\,Mb/s data rate threshold respectively. Compared with the result in Fig.~\ref{fig:RIS-free-optimal-random} (a), the minimum SNR that one UE holds under optimal phase shift is 18\,dB, and the coverage is clearly extended. And the minimum SNR among all UEs is still lower than $-10$\,dB when a random phase shift is applied. Benefiting from the capability to control the wireless environment, RIS with the optimal phase shift can obviously improve the SNR condition. 
    
        \vspace{-1mm}
\section{Conclusion}
In this paper, we propose an RIS deployment algorithm in IDS to extend the coverage of the network. As a case study, we first integrate RIS into RT simulations considering an airplane cabin, and provide guidelines to realistically capture the signal propagation to and from an RIS. We then formulate and solve the RIS deployment problem that jointly minimizes the number of RISs, selects the RIS phase-shifts, and allocates time resources to the UEs to satisfy given data rate thresholds. We observe that deploying at least $4$ RISs can improve the coverage from $3$ rows to $11$ rows, thereby providing at least a $18$\,dB received power gain to the UEs previously in the outage. With $17$ RISs having $256$ elements, each UE can get $130$\,Mbit/s  data rate. %As future work, we aim to extend the RIS deployment to the whole cabin, and investigate the performance with multi-user MIMO transmission. 

%\balance

\section*{Acknowledgement}
Results incorporated in this paper received funding from the ECSEL Joint Undertaking (JU) under grant
agreement No 876124. The JU receives support from the EU Horizon 2020 research and
innovation programme and Vinnova in Sweden.

\bibliographystyle{IEEEtran}

\bibliography{InCabin}

\end{document}